# Observation of Fundamental Thermal Noise in Optical Fibers down to Infrasonic Frequencies

Jing Dong[1, 2], Junchao Huang[1, 2], Tang Li[1, a)], and Liang Liu[1, a)]

[1]*Key Laboratory of Quantum Optics and Center of Cold Atom Physics, Shanghai Institute of Optics and Fine Mechanics, Chinese Academy of Sciences, Shanghai 201800, People's Republic of China*

[2]*University of Chinese Academy of Sciences, Beijing 100049, People's Republic of China*

The intrinsic thermal noise in optical fibers is the ultimate limit of fiber-based systems. However, at infrasonic frequencies, the spectral behavior of the intrinsic thermal noise remains unclear so far. We present the measurements of the fundamental thermal noise in optical fibers obtained using a balanced fiber Michelson interferometer. When an ultra-stable laser is used as the laser source and other noise sources are carefully controlled, the 1/f spectral density of thermal noise is observed down to infrasonic frequencies and the measured magnitude is consistent with the theoretical predictions at the frequencies from 0.2 Hz to 20 kHz. Moreover, as observed in the experiment, the level of 1/f thermal noise is reduced by changing the coating of optical fibers. Therefore, a possible way to reduce the thermal noise in optical fibers at low Fourier frequencies is indicated. Finally, the inconsistency between the experimental data on thermomechanical noise and existing theory is discussed.

In many fiber-based systems, such as interferometric fiber-optic sensors[1-4], fiber lasers[5-7], and fiber-delay-line stabilized lasers[8], the fundamental resolution or frequency stability is limited by the intrinsic thermal noise inside optical fibers. Thus, an experimental and theoretical understanding of the thermal noise in optical fibers is important.

Over the past three decades, such thermal noise has been investigated theoretically[9-18] and experimentally[1, 7, 19-23]. Glenn[9] first calculated the fluctuations of thermodynamic phase in optical medium and showed that they can be larger than shot noise over a wide frequency range. In the early 1990s, Wanser[10] proposed a brief formula for estimating the magnitude of the thermal noise in optical fibers. Subsequently, a similar but more detailed model was presented by Foster *et al.*[14] in the context of distributed feedback (DFB) fiber lasers. According to Wanser's theory[10], in a fiber, the power spectral density (PSD) of the phase noise originating from thermodynamic contribution is given by

$$S_T(\omega) = \frac{2\pi k_B T^2 L}{\lambda^2 \kappa}\left(\frac{dn}{dT}+n\alpha_L\right)^2 \ln\left(\frac{k_{max}^4 + (\frac{\omega}{D})^2}{k_{min}^4 + (\frac{\omega}{D})^2}\right) \quad (1)$$

where $k_B$ is the Boltzmann constant, $T$ is the temperature, $L$ is the fiber length, $\lambda$ is the wavelength, $dn/dT$ is the refractive index temperature coefficient, $n$ is the effective refractive index, $\alpha_L$ is the thermal expansion coefficient, $\kappa$ is the thermal conductivity, $D$ is the thermal diffusivity, $k_{max}$ and $k_{min}$ are the boundary conditions. This expression is widely used for estimating the magnitude of the thermal noise in optical fibers and is consistent with the experimental results at high Fourier frequencies[1, 22, 23]. However, in the low frequency range, the theory predicts a nearly frequency-independent spectrum of thermal noise, which is not consistent with the experiment data[23].

Recently, spontaneous fluctuations of fiber length induced by mechanical dissipation were proposed by Duan[15, 16] as another possible source of the thermal noise in optical fibers. The model is based on the fluctuation-dissipation theorem (FDT) and shows that the fluctuations of fiber length (the so-called thermomechanical noise by Duan that differs from the thermoconductive noise proposed by Wanser and Foster *et al.*), representing the phase noise in fiber interferometers, have a $1/f$ spectral density. At low frequencies, the phase noise PSD of this contribution in a fiber can be described by[15]

$$S_M(f) = \left(\frac{2\pi}{\lambda}n\right)^2 \frac{2k_B TL\phi_0}{3\pi E_0 A}\frac{1}{f} \quad (2)$$

where $E_0$ is the bulk modulus of the material, $A$ is the cross-sectional area of the fiber, and $\phi_0$ is the loss angle that characterizes mechanical dissipation. A later experiment performed by Batolo *et al.*[23] shown that the combination of thermomechanical and thermoconductive models is consistent with the experimental data extracted from an optical fiber interferometer at the frequencies from 30 Hz to 100 kHz. For the range of infrasonic frequencies (10 mHz to 30 Hz), Gagliardi *et al.*[4] reported strain measurements with the highest sensitivity obtained by using a fiber Fabry-Perot (FP) cavity. However, the sources of the noise floor are still in doubt[18, 24, 25]. Recently, in the experiment for ultra-stable fiber-stabilized laser, Dong *et al.*[8] observed that the frequency

---
a) Electronic mail: litang@siom.ac.cn, liang.liu@siom.ac.cn

noise PSD range from 1 Hz to 100Hz exhibits a $1/f$ characteristic seems consistent with Duan's theory.

In spite of the recent experimental progress, the thermal noise below 30 Hz is still unclear owing to the excess noise sources such as laser noise, electronic noise and environmental noise. In this paper, the fundamental thermal noise in optical fibers is investigated down to infrasonic frequencies using an optical fiber interferometer. The measured noise level is consistent with the theoretical predictions at the frequencies from 0.2 Hz to 20 kHz. Furthermore, by using optical fibers with different coating, we studied the relationship between thermomechanical noise and fiber mechanical properties. This study helped to find a possible way to reduce the level of the thermal noise in optical fibers in the low frequency range. In order to circumvent the problem of the noise sources mentioned above, we adopt a heterodyne detection scheme and use a sub-hertz linewidth laser as the laser source which reduce the baseband noise at low Fourier frequencies dramatically. Additionally, the minimization of thermal drift by using a sophisticated environmental isolation and active compensation allows measuring the thermal noise at frequency down to 0.01 Hz.

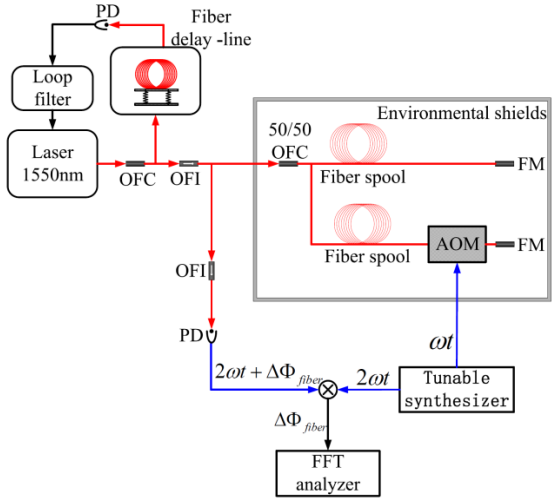

FIG. 1. Experimental setup of the measurement of the thermal noise in optical fibers. AOM (acousto-optical-modulator), FM (Faraday-mirror), PD (photodiode), OFC (optical fiber coupler), OFI (optical fiber isolator), FFT (fast Fourier transform).

We demonstrated the measurements of thermal noise using a sub-hertz laser to interrogate balanced fiber Michelson interferometer (BFMI) (FIG.1). An acousto-optical modulator (AOM) was inserted into one arm after a fiber delay line and was driven by 75 MHz to shift the optical frequency for heterodyne detection, which minimizes the baseband noise from laser intensity and detection system. In the interferometer, two Faraday mirrors were used to guarantee the maximum amplitude of beat note signal without the requirement of any polarization controller. The optical power that injected into the interferometer was approximately 8 mW. At photodiode, a RF signal of 150 MHz was detected. A low phase noise tunable synthesizer fabricated by us was used to provide the demodulation signal. The demodulated signal was sent to a FFT spectrum analyzer (SR760, Stanford Research Systems) to extract the phase noise in optical fibers. To reduce the impact of laser frequency noise, a sub-hertz laser was used as the laser source. The laser was stabilized onto an optical fiber delay-line and its linewidth is narrower than 0.67 Hz[8]. In order to minimize the differential environmental pickup in both arms of the interferometer, the optical fibers were co-wound onto a solid titanium cylinder under very low tension (approximately 0.1 N). The interferometer was thermally shielded and placed into a vacuum tank while the vacuum tank was placed on a passive vibration isolation platform (BM-4, Minus K Inc., its lowest cutoff frequency was 0.5 Hz) and housed by an aluminum box. The aluminum box was covered with acoustic-damping foam and thermal insulation foam to isolate the environmental noises. Next, the temperature was stabilized at approximately 300 K by an active thermal controller. The total temperature fluctuation was less than 10 mK over a period of 24 hours. Furthermore, by applying a small frequency offset to the tunable synthesizer, the linear thermal drift could be removed after the long-term temperature stabilization. Therefore, a stable quadrature measurement of phase noise could be performed.

Table I. Summary of the fiber parameters used for theoretical calculation

| Property | Notation | Value |
|---|---|---|
| Wavelength | $\lambda$ | 1550 nm |
| Temperature | $T$ | 300 K |
| Refractive index temperature coefficient | $dn/dT$ | $9.2\times10^{-6}$ /K |
| Effective refractive index | $n$ | 1.468 |
| Thermal expansion coefficient | $\alpha_L$ | $5.5\times10^{-7}$ /K |
| Thermal conductivity | $\kappa$ | 1.37 W/(mK) |
| Thermal diffusivity | $D$ | $8.2\times10^{-7}$ m$^2$/s |
| Boundary condition parameter | $k_{max}$ | $3.846\times10^{5}$ /m |
| Boundary condition parameter | $k_{min}$ | $3.848\times10^{4}$ /m |
| Bulk modulus of the material | $E_0$ | $1.9\times10^{10}$ Pa |
| Cross-sectional area of the fiber | $A$ | $4.91\times10^{-8}$ m$^2$ |
| Loss angle | $\phi_0$ | 0.01 |

The BFMI is constructed from Corning standard SMF-28 optical fibers (9/125/250) with a length of 327 m in each arm. For a Michelson scheme, the total length of the fiber, which induces phase noise, should be 1308 m. The length mismatch is less than 1 m and, owing to the ultra-stable laser source, the phase noise induced by the fluctuations of laser frequency can be neglected (see curve (c) in FIG. 2). FIG. 2 shows the results of phase noise measurement compared with the theoretical predictions and other contributing noise sources. Obviously, the noise floor from tunable synthesizer (curve (b)) and the shot noise floor from photodetction process (curve (d)) are more than 15 dB lower than the measured phase noise of optical fibers and, thus, can be neglected. Curve (e) is the combination of both Duan's theory and Wanser's theory calculated based on the fiber parameters in Table I. Curve (e) shows a small offset from the experimental data at low frequencies. To remove the offset, a small factor of 2.5dB is subtracted from the calculations by Duan's theory. This subtraction is reasonable because the values of some fiber parameters used for calculations such as the loss angle $\phi_0$ and the bulk modulus $E_0$ have not yet been verified experimentally at low frequencies[26]. It can be observed that the experimental data are consistent with the theoretical predictions (curve (f)) in the frequency range from 0.2 Hz to 20 kHz. The phase noise is dominated by very slow temperature fluctuations at the frequencies lower than 0.2 Hz and by the residual laser noise bump at the frequencies higher than 20 kHz.

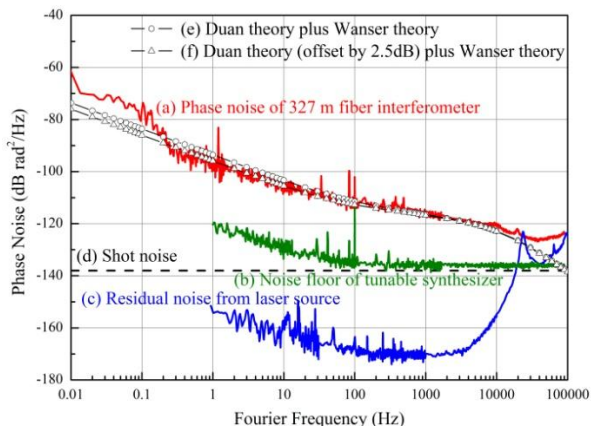

FIG. 2. Phase noise PSD of the optical fiber interferometer due to the thermal noise and contributing noise sources. (a), Measured phase noise of 327 m fiber interferometer (red solid line). (b), Noise floor of tunable synthesizer (green solid line). (c), Residual phase noise from laser source due to arm imbalance (blue solid line). (d), Calculated shot noise floor (black dashed line). (e), Combination of Duan's theory and Wanser's theory (circles). (f), Combination of Duan's theory subtracted by 2.5 dB and Wanser's theory (triangles).

According to the theoretical predictions, the magnitude of thermal noise is proportional to the fiber length. A comparison of the phase noise of four different fiber lengths, 327 m, 134 m, 68 m and 34 m BFMIs, respectively, is shown in FIG. 3. One would expect a 134 m, 68 m, and 34 m BFMIs to have a 3.9 dB, 6.8 dB and 9.8 dB smaller thermal noise relative to a 327 m BFMI, respectively. As can be clearly seen, the experimental data are consistent with the scaling of the thermal noise for different fiber lengths.

According to the expression (2), the PSD of thermomechanical noise is proportional to the loss angle $\phi_0$. Therefore, it implies that the level of thermomechanical noise can be suppressed by means of the reduction of the loss angle of fiber. Because the Q factor of fused-sillica is higher than $1 \times 10^6$ [27], the major contribution to the loss angle of optical fibers comes from the polymer coating. This provides a possibility to change the values of the loss angle by varying the fiber coating. Furthermore, according to the lumped element model proposed by Beadle et al.[26], the variation of the fiber coating does not change the product $E_0 \times A$ significantly because the Young's modulus of the silica glass core of fiber is much higher than that of polymer coating. Consequently, one would expect a lower level of thermomechanical noise in optical fibers with a thinner polymer coating. In order to verify this assumption, we repeated the measurements of phase noise using a 103 m BFMI constructed from Yangtze HT1510-A fibers (9/125/155), which had a 15-μm-thick polyimide (PI) coating. The result is shown in FIG. 4. For a direct comparison, the phase noise of a 134 m SMF-28 BFMI, which is subtracted by a scaling factor of fiber length of 1.14 dB, is also presented in FIG. 4. Obviously, at low Fourier frequencies, the phase noise PSD of PI coating fibers is $2.5 \times 10^{-11} \times f^{-1}$ $rad^2/Hz$ which is approximately 5 dB lower than that of SMF-28 fibers. It means that, if this PI coating BFMI was used as a 1550 nm strain sensor, a strain resolution of approximately $5.2 \times 10^{-15} \varepsilon/\sqrt{Hz}$ ($\varepsilon$ is fractional length change) at 1 Hz is achievable. Although it is very difficult to constitute a quantitative relationship between the thermomechanical noise and the properties of fiber coating because of the lack of the experimental results on $E_0$ and $\phi_0$ for both SMF-28 fibers and PI coating fibers, these results indicate a possible way to reduce the thermal noise in optical fibers at low Fourier frequencies.

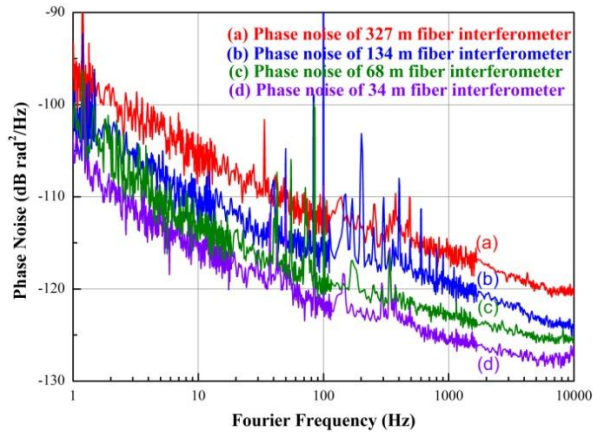

FIG. 3. Measured phase noise PSD of (a) a 327 m fiber interferometer (red line), (b) 134 m fiber interferometer (blue line), (c) 68 m fiber interferometer (green line) and (d) 34 m fiber interferometer (violet line).

Although the theory and experimental data are in good agreement on the magnitude and spectral shape of thermal noise, the resonance peaks of the thermomechanical noise predicted by Duan were not observed in the experiment. According to Duan's theory[16], the spontaneous length fluctuations of optical fibers can be analyzed via normal mode expansion and each normal mode can be treated as a harmonic oscillator, which naturally leads to the presence of resonance peaks in the spectrum of thermal noise. The resonant frequencies are proportional to the inverse of fiber length. In the experiment, we measured thermal noise with 327 m, 134 m, 103 m, 68 m and 34 m BFMIs, respectively, and, thus, the resonant frequencies of different BFMIs should be 5.6 Hz, 13.6 Hz, 17.8 Hz, 26.8 Hz, and 53.6 Hz, and their harmonics, respectively. However, we do not observe any regular resonance peaks in the experimental curves. The satellite resonance peaks in the spectrum around 1 Hz and 40 Hz are due to the seismic noise and acoustic noise, respectively. The peaks distributed at the frequencies from 80 Hz to 1 kHz come from the pickup of power ground noise. In Duan's theoretical publication[16], he attributed such phenomenon to a higher loss angle. According to the experimental data, this assumption is not correct since the loss angle derived from the measured thermal noise is even lower than 0.01. Moreover, at low Fourier frequencies, the resonance peaks should be clearly observed even at a loss angle of 0.1 because the thermomechanical noise dominates the frequency range below 100 Hz. Consequently, this inconsistency between the experimental data and the model proposed by Duan indicates, perhaps, a new physical mechanism needed to be further investigated.

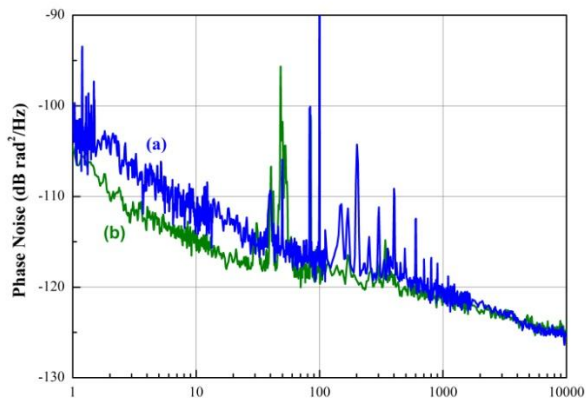

FIG. 4. Measured phase noise PSD of (a) a 134 m SMF-28 fiber interferometer subtracted by a factor of 1.14 dB (blue line, upper trace) and (b) 103 m PI coating fiber interferometer (green line).

In conclusion, we observed the effects of the phase noise induced by the fundamental thermal noise in optical fibers at the frequencies from 0.2 Hz to 20 kHz based on the heterodyne balanced fiber Michelson interferometer. The measured noise levels are consistent with the theoretical predictions. By changing the coating of optical fibers, we demonstrated a reduction of the level of $1/f$ noise which helps to find a possible way to improve the resolution of fiber-based systems, such as fiber-optic sensors and fiber interferometers, in the infrasonic frequency range. In addition, the mechanism and theoretical model of the thermal noise in optical fibers in the low frequency range require further investigations in future.

This research is supported by the National Natural Science Foundation of China under Grant Nos. 11034008 and 11274324 and the Key Research Program of the Chinese Academy of Sciences under Grant No. KJZD-EW-W02.